\documentclass[aps,prb,twocolumn,superscriptaddress]{revtex4-2}
\usepackage{amssymb,amsmath}
\usepackage[version=4]{mhchem} %For nice and easy chemical formulas, like \ce{H20} (instead of H$_2$O)
\usepackage{graphicx}% Include figure files
\usepackage{epstopdf}
\usepackage{physics} %Useful definitions of vectors, brakets etc.
\usepackage[pdfstartview=FitH,breaklinks=true,bookmarks=true,colorlinks=true,anchorcolor=black,citecolor=red,filecolor=black,menucolor=black,urlcolor=blue,linkcolor=blue]{hyperref}

\begin{document}

\title{Frustration induced incommensurate solids in the extended Bose-Hubbard model}

\author{Kwai-Kong Ng}
\email[]{kkng@thu.edu.tw}
\affiliation{Department of Applied Physics, Tunghai University, Taichung 40704, Taiwan (R.O.C.)}

\author{Min-Fong Yang}
\email[]{mfyang@thu.edu.tw}
\affiliation{Department of Applied Physics, Tunghai University, Taichung 40704, Taiwan (R.O.C.)}

\date{\today}

\begin{abstract}
The extended Bose-Hubbard model with nearest-neighbor and next-nearest-neighbor repulsive interactions on a square lattice is investigated by using the quantum Monte Carlo method. We find that, for the cases of weak next-nearest-neighbor interactions and small hoppings, incommensurate solids of fractional densities varying from 1/4 to 1/2 can be stabilized in the thermodynamic limit. We further show that the continuous changes of ordering wave vectors from $(\pi,\pi/2)$ [or $(\pi/2,\pi)$] to $(\pi,\pi)$ within the incommensurate solid phase can be understood by the mechanism of domain wall formation. The related ground-state phase diagram and thermal phase transitions are also discussed.
\end{abstract}

% insert suggested PACS numbers in braces on next line
% insert suggested keywords - APS authors don't need to do this
%\keywords{}

\maketitle

\section{introduction}

Polar molecules trapped in optical lattices provide an unique opportunity to study the dipole-dipole interactions in real experimental setups \cite{Lahaye2009,Ni231,Ospelkaus622,Landig2016}. One primary investigation is to identify possible quantum phases that may uniquely arise from these long-range interactions but cannot be observed in systems with only short-range interactions. Recent numerical studies \cite{PhysRevLett.104.125301,PhysRevA.86.063635,PhysRevB.86.054516,PhysRevA.101.013616} of hard-core bosons with infinite-range interactions on two-dimensional (2D) square lattices have presented evidence of Mott insulators with checkerboard, stripe, and star ordering at densities $\rho=$ 1/2, 1/3, and 1/4, respectively. Furthermore, supersolids around the Mott lobes with $\rho=$ 1/2 and 1/4 are also found by doping the solids with particles or vacancies. Besides, devil's staircase (that is, a sequence of commensurate phases separated by a series of jumps) is observed for finite-size systems, which signals the presence of the incommensurate phases in the thermodynamic limit. A question is: Are all these interesting phases uniquely stabilized by long-range interactions?

Actually, the Mott insulators and the supersolids can be realized in short-range models as well. In previous numerical investigations on the extended Bose-Hubbard model that includes only the nearest-neighbor (nn) and the next-nearest-neighbor (nnn) interactions, Mott insulators at 1/2 and 1/4 fillings and the associated supersolids have indeed been observed \cite{PhysRevLett.84.1599,PhysRevB.65.014513,Ng_2010,PhysRevB.77.052506,PhysRevB.78.132512,PhysRevB.82.184505}. The model Hamiltonian for the hard-core bosons on a 2D square lattice is,
\begin{align}
  H=&-t\sum_{\langle i,j \rangle}( b_i^\dagger b_j +h.c.) + V \sum_{\langle i, j \rangle} n_i n_j + V'\sum_{\langle i, j \rangle '} n_i n_j  \nonumber \\
  &- \mu \sum_i n_i \; .
  \label{eq_xBH}
\end{align}
Here $b_i$ ($b^\dagger_i$) is the annihilation (creation) bosonic operator on site $i$, $t$ the hopping integral, $n_i$ the particle number on site $i$, and $V$ ($V'$) the nn (nnn) repulsive interaction. $\mu$ denotes the chemical potential which controls the particle density in the grand canonical ensemble. For the cases of strong $V'$ ($V'>V/2$), Mott insulators of striped and star orders are observed at 1/2 and 1/4 fillings respectively. Both of them are associated with supersolids of the same orderings. Therefore, the long-range nature of the dipole-dipole interactions is not indispensable to the stability of these phases.

Moreover, the discovery of incommensurate supersolid phases for hard-core bosons with nn interactions on anisotropic triangular lattices~\cite{Isakov_2009,PhysRevLett.117.193201} shows that long-range interactions are not crucial for the appearance of incommensurate phases. This motivates us to explore the possibility of incommensurate phases in short-range models. In addition, the nature of incommensurate phases and their stabilities is of great interest on its own, the understanding of which should shed light on the other systems with frustrated interactions. However, these fundamental issues are hard to address for infinite-range models. The realization of incommensurate phases in short-range models will thus help us to analyze the mechanism of their formation and then to improve our understanding of these phases.

In this work, we employ the quantum Monte Carlo (QMC) method with a stochastic series expansion algorithm \cite{PhysRevB.59.R14157,PhysRevE.66.046701} to study the short-range model in Eq.~\eqref{eq_xBH}. We find that, at small hopping $t$ and weak nnn interactions $V'$ ($V'<V/2$), incommensurate solids of densities ranging from 1/4 to 1/2 can be stabilized in the thermodynamic limit. The ground-state phase diagram, as one of our main results, is presented in Fig.~\ref{fig1}, where incommensurate solids emerge between the half-filled checkerboard and quarter-filled solids. In contrast to the strong $V'$ cases \cite{PhysRevLett.84.1599,PhysRevB.65.014513,PhysRevB.77.052506,PhysRevB.78.132512,Ng_2010}, there exists no supersolid phases in the present case of weak $V'$. The incommensurate solids are found to be characterized by a continuous change in ordering wave vectors as model parameters are varied. The formation of such a phase can be explained by the proliferation of domain walls from half-filled or quarter-filled solids. Following the analysis in Ref.~\cite{PhysRevLett.117.193201} in the context of the anisotropic triangular lattice, analytical calculations on the domain-wall excitation energies are made for the present system on a square lattice. Our analytical predictions on the phase boundaries agree very well with the QMC numerical results. Furthermore, the quantum phase transitions from the incommensurate solids to the half-filled or quarter-filled solids are found to be continuous in the thermodynamic limit, while the transition to the superfluid phase is first order. Upon increasing temperatures, the incommensurate solids will melt into normal fluids via a first-order thermal phase transition, which is demonstrated by hysteresis of the structure factors and double peaks in the histogram.

This paper is organized as follows. The ground-state phase diagram at fixed $V$ and $V'$ is identified in Sec.~II. The general features of the incommensurate solids are explained in Sec.~III by the domain wall formation. The energetic perspective of the domain wall is elucidated in Sec.~IV. Analytic expressions of phase boundaries of the incommensurate solids and of the domain wall density are then derived. More quantum and thermal phase transitions out of the incommensurate solids are discussed in Sec.~V. We summarize our work in Sec.~VI.

\section{Quantum phase diagram}
\label{QPD}
In this work we utilized the well established stochastic series expansion algorithm \cite{PhysRevB.59.R14157,PhysRevE.66.046701} in the QMC calculations as the Hamiltonian $H$ in Eq.~\eqref{eq_xBH} is sign-problem free. Unless mentioned otherwise, we set $V'=1$ as the energy scale. The inverse temperature is set to be $\beta=1/(2L)$, where $L$ is the linear size of lattice ($L=12$, 24, 36, and 48 are used). To identify various quantum phases, besides the particle density $\rho$, two more quantities are measured in our simulations., The superfluidity $\rho_s$ is calculated to signal off-diagonal U(1) symmetry breaking, which is defined by the fluctuations of winding numbers $W_x$ and $W_y$,
\begin{equation}
\rho_s = \frac{1}{2\beta} (\langle W_x^2 \rangle + \langle W_y^2 \rangle) .
 \label{rho_s}
\end{equation}
On the other hand, the translational broken symmetry is characterized by the structure factor
\begin{equation}
S(\textbf{Q}) = \frac{1}{L^2} \sum_{i,j} \langle n_i n_j \rangle e^{i \textbf{Q} \cdot \textbf{r}_{ij}},
  \label{struct_ftr}
\end{equation}
where \textbf{Q} denotes the wave vector and \textbf{r}$_{ij}$ the displacement between two sites $i$ and $j$.

\begin{figure}
\includegraphics[width=\columnwidth]{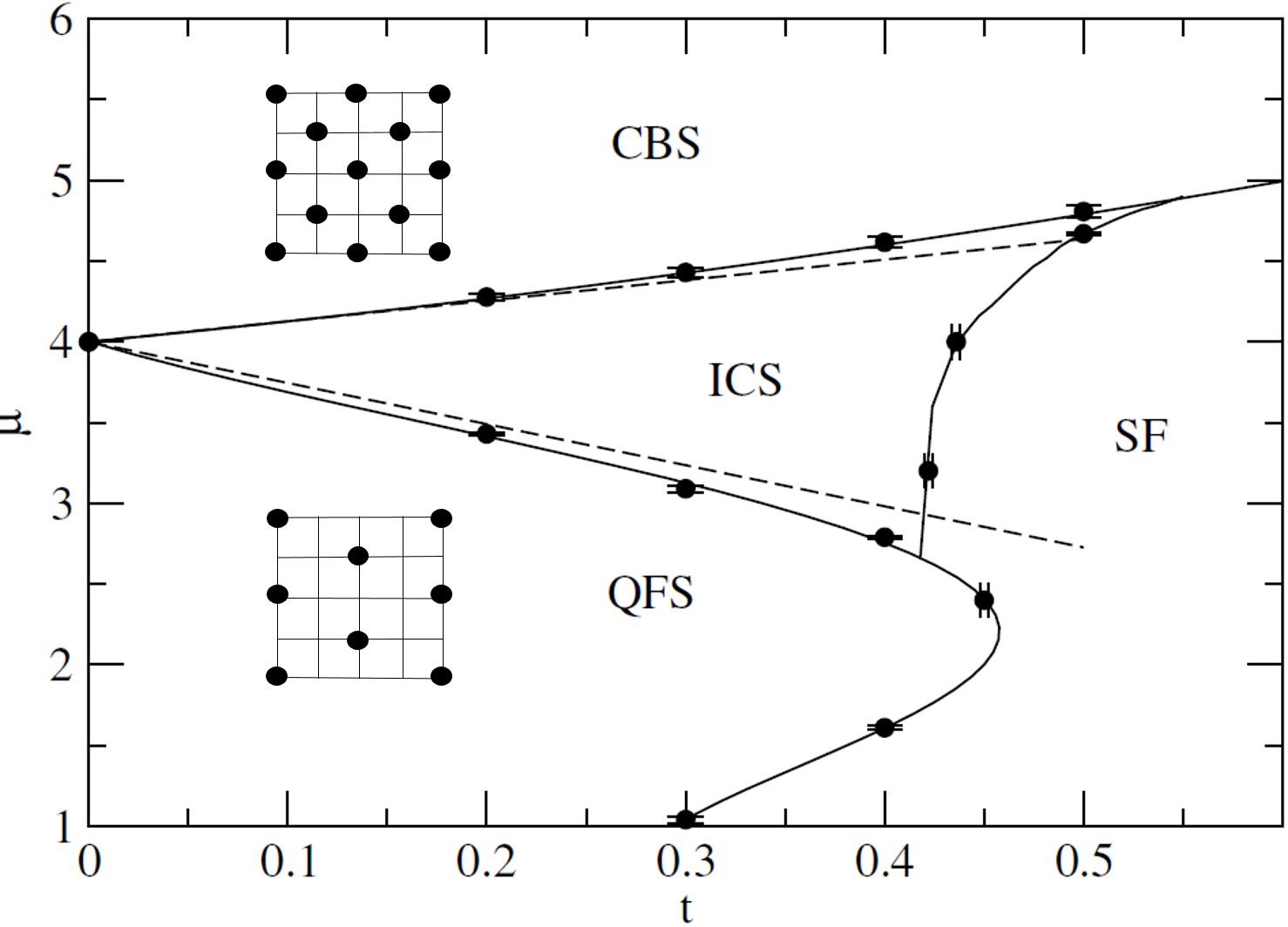}
\caption{
Ground-state phase diagram %$\mu$ vs $t$
with $V=4$ obtained from QMC calculations after extrapolating to thermodynamic limit. The dotted lines are the phase boundaries predicted by the domain wall analysis. The insets show the CBS and the QFS lattice structures. Here $V'=1$ as the energy unit.
\label{fig1}}
\end{figure}

In the parameter regime we studied in our model, the off-diagonal and the diagonal orderings do not coexist, which means no supersolid is found.
In Fig. \ref{fig1}, the ground-state phase diagram $\mu$ vs $t$ with $V=4$ contains a superfluid (SF) and two solid states of definite ordering wave vectors \textbf{Q} signaling the peak in the structure factor. Their phase boundaries are determined by the abrupt changes of $\rho_s$ or $S(\textbf{Q})$ (see Fig. \ref{fig2}) and then by extrapolating to the thermodynamic limit with $L \rightarrow \infty$. The half-filled phase is a checkerboard solid (CBS) with $\textbf{Q}=(\pi,\pi)$, instead of a striped solid, because the nn repulsion is dominant ($V >2V'$) so that nn occupation is avoided. Reducing the chemical potential such that $\rho =1/4$, the ground state is then a quarter-filled solid (QFS) of star ordering \cite{PhysRevB.77.052506,PhysRevB.78.132512} with $\textbf{Q}=(\pi/2,\pi)$ or $(\pi,\pi/2)$. While there are two types of star orders that are doubly degenerate at $t=0$, the degeneracy is lifted for finite $t$.  The preferred QFS ordering is shown in the inset of Fig. \ref{fig1} as it prevents nn occupation up to second order of hopping $t$ to reduce the cost of the nn potential energy.  Interestingly, for finite but small $t$, the phase in between the CBS and the QFS is neither a superfluid nor supersolid, but an incommensurate solid (ICS) with ordering wave vectors \textbf{Q} continuously varying from $(\pi/2, \pi)$ [or $(\pi, \pi/2)$], to $(\pi,\pi)$ in the thermodynamic limit. As explained in the next section, we found that ICS can be understood by inserting linear domain walls into the solid phases. By increasing hopping $t$, all solid phases eventually melt into a superfluid phase directly without passing an intermediate supersolid phase. Further discussions on these phase transitions will be provided in Sec. \ref{QPT}.

\section{incommensurate solids}
\label{ICS}
\begin{figure}
\includegraphics[width=\columnwidth]{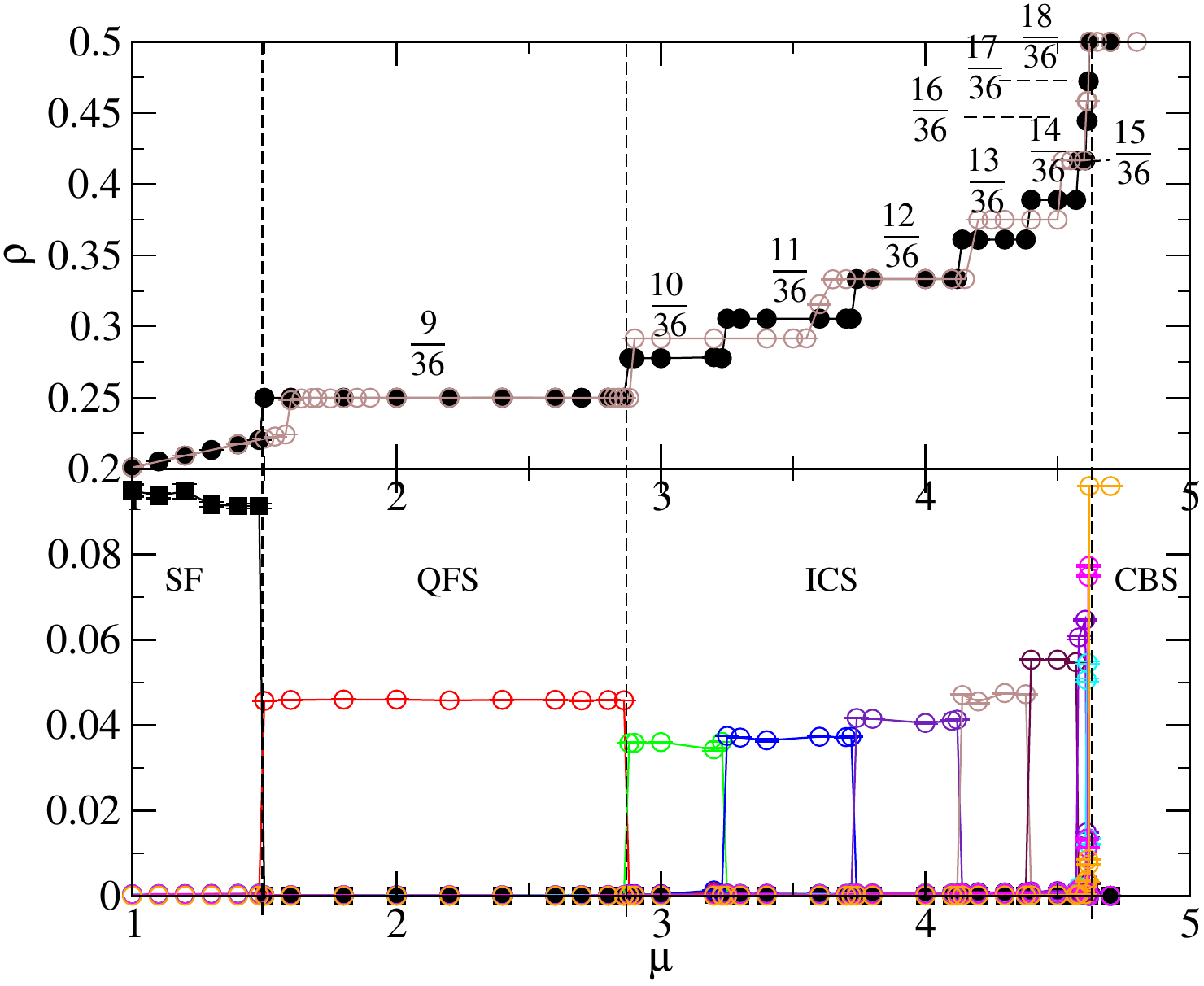}
\caption{
Order parameters as functions of $\mu$ with $t=0.4$ and $V=4$. Upper panel: the particle density $\rho$  as a function of $\mu$. The fractional numbers are the corresponding densities of the plateaus. Solid (Open) circles represent the data for $L=36$ ($L=24$). Lower panel: superfluidity $\rho_s$ (filled squares) and structure factors $S(\textbf{Q})$ (open circles) as functions of $\mu$. The corresponding ordering wave vectors of the data are, from left to right,  \textbf{Q}=($2m\pi/L$, $\pi$) with $m=L/4$, $(L/4)+1$, ..., $L/2$. For simplicity, the structure factors for the ordering wave vectors \textbf{Q}=($\pi$, $2m\pi/L$) are not shown. Note that the data of $S(\pi,\pi)$ is reduced by five times.
\label{fig2}}
\end{figure}

For a given hopping $t=0.4$, the particle density $\rho$ is shown in Fig. \ref{fig2} for $L=36$. A series of density plateaus is observed from $\rho=1/4$  to 1/2, in between of the QFS and the CBS. We note that the plateaus occur at densities of $m/L$, with $m$ being an integer from $L/4$ to $L/2$. This is very similar to the results found in the model with infinite-range interactions \cite{Ospelkaus622,PhysRevLett.104.125301}. In that case, although a smaller lattice size ($L=12$) is used, many more plateaus of rational fillings are found between $\rho$ =1/4 and 1/2.
The reason for this difference is that the infinite-range nature of the interactions allows more possible configurations of particle ordering to be stabilized, which, unfortunately, also makes the physics more difficult to be analyzed. In contrast, for the present system with short-range interactions, we can understand the origin of the density plateaus by a simple picture of linear fluctuating domain walls.
While both domain walls along the $x$ and $y$ directions are observed in our simulations, for simplicity, only the domain walls along the $y$ direction will be discussed hereafter.

Under doping the half-filled CBS with holes, the potential energy gain is $4V'$ for each added isolated hole. Nevertheless, holes aligned together to form a domain wall [Fig. \ref{fig3} (b)] can further gain kinetic energy through particle hoppings, which will be explained in details in the next section when we discuss the domain wall dynamics. For convenience, we define $L_x$ ($L_y$) to be the lattice size in the $x$ ($y$) direction. We note that even numbers of domain walls are required to maintain the periodic boundary condition imposed in our system. Because $L_y/2$ particles must be removed in order to create a domain wall along the $y$ direction out of the CBS, the total number of removed particle becomes $nL_y$ when $2n$ domain walls appear in the lattice for $n=0$, 1, 2, ..., $L_x/4$. Consequently, when there are $L_x/4$ pairs of domain walls, the particle density is reduced by $1/4$ and the state becomes the QFS with $\rho=1/2-1/4=1/4$ [Fig. \ref{fig3} (d)]. For the states containing $2n$ domain walls, the particle densities become $\rho=(L_x-2n)/(2L_x)$, which exactly correspond to those of the plateaus found in Fig.~\ref{fig2} for both $L=24$ and 36 (note that $L=L_x=L_y$ in all our simulations). Our analysis clearly shows that the series of plateaus is formed by adding pairs of linear domain walls from the CBS.

Furthermore, the observed shift in the ordering wave vector \textbf{Q} in Fig.~\ref{fig2} implies the presence of linear domain walls as well. We note that the insertion of domain walls at equal distance will not immediately destroy the order completely. Instead, due to producing long-period superstructures in the $x$ direction, the ordering wave vector will be shifted by ($2\pi/L_x, 0$) for each added pair of linear domain walls along the $y$ direction. Hence the plateau states in ICS with $2n$ domain walls along the $y$ direction can be characterized by the ordering wave vectors $\textbf{Q}=(\pi-2n\pi/L_x,\pi)$ or $\textbf{Q}=(2m\pi/L_x,\pi)$ with $m$ being an integer from $L_x/4$ to $L_x/2$. (The same discussions apply as well for the states with $2n$ domain walls along the $x$ direction.) This conclusion is consistent with the numerical results shown in Fig~\ref{fig2}.

Another implication from the above analysis is that as $L \rightarrow \infty$, the number of plateaus will become infinity such that $\rho$ changes continuously from 1/4 to 1/2. The density jump at the phase boundary between QFS (CBS) and ICS is then expected to be reduced to zero and the corresponding commensurate-incommensurate transition should thus be continuous, rather than of first order. The location of these phase boundaries in the ground-state phase diagram for small $t$ can be determined analytically by considering the domain wall dynamics, as discussed in the next section.

\section{domain wall dynamics}
\label{DW}
\begin{figure}
\includegraphics[width=7cm]{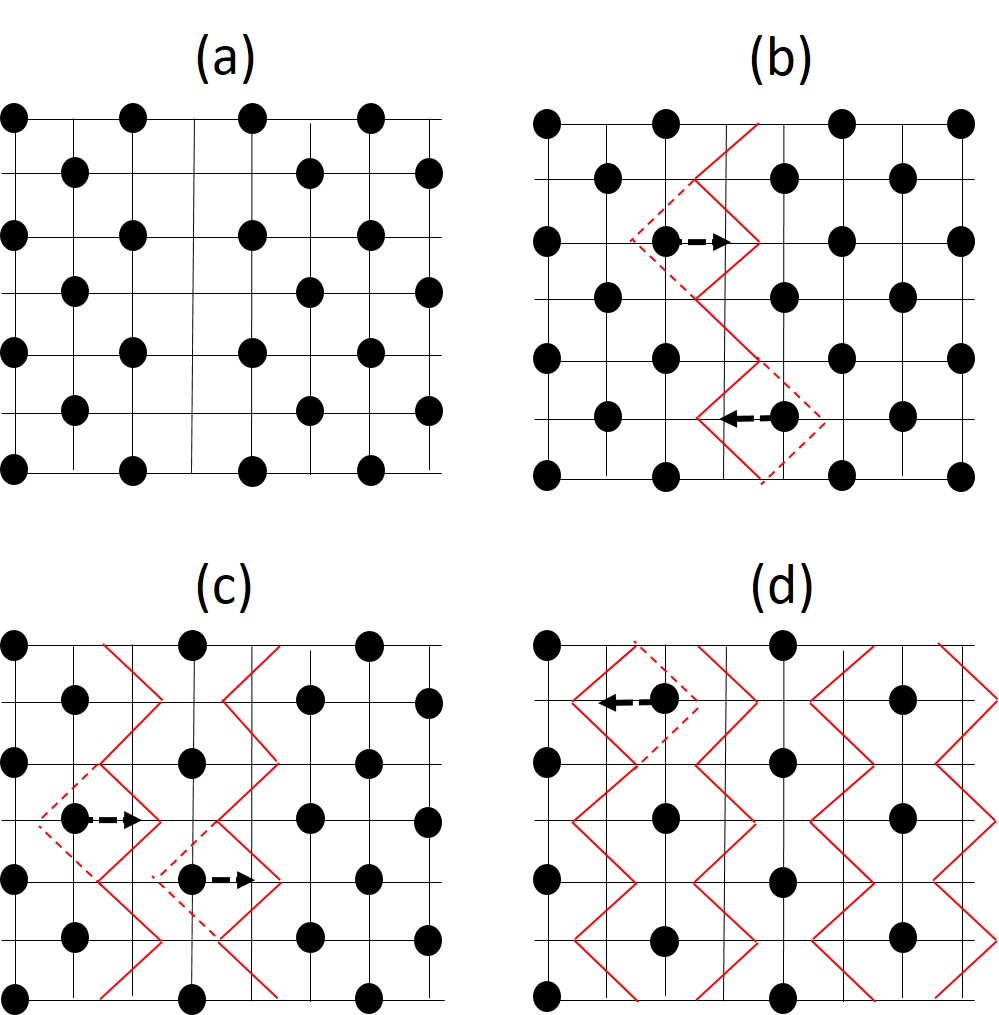}
\caption{
Illustrations of domain wall motion. (a) Removing a column of bosons. (b) Shifting the right half of the lattice costs no extra potential energy but gains kinetic energy due to the fluctuation of the domain wall (red line). (c) Multiple domain walls. (d) QFS state with a domain wall density $\rho_D = 1/2$.
\label{fig3}}
\end{figure}

Here we follow the analysis in Ref.~\cite{PhysRevLett.117.193201} for incommensurate supersolids on anisotropic triangular lattices to discuss the domain wall dynamics in the present model. Started from the half-filled CBS state, one can split the CBS into half by removing a column of bosons (i.e., removing $L_y/2$ bosons), as depicted in Fig. \ref{fig3}(a). This gives a cost in potential energy per unit length $\Delta E_p /L_y = (\mu-4V')/2$. After shifting half of the lattice upwards, a domain wall is created without further cost in potential energy. To be specific, we define the domain walls by connecting the center of unoccupied bonds, shown as the red solid zigzag line in Fig. \ref{fig3}(b). Note that the domain wall can fluctuate to gain the kinetic energy, because the bosons around the shifted boundary are now free to move sideways. As illustrated in Fig. \ref{fig3}(b), after the hopping of bosons, parts of the domain wall fluctuate in the opposite direction of the particle movement (shown as the red dotted lines). To estimate the gain in kinetic energy, one can map the one-dimensional (1D) domain wall onto a 1D spin-1/2 XY chain \cite{PhysRevLett.117.193201,PhysRevB.49.14047,PhysRevB.68.014506}. That is, by tracing the zigzag chain along the $y$ direction, a up (down) spin is assigned if the $x$-coordinate of a given unoccupied bond is increased (decreased) by one unit. The kinetic energy of a domain wall thus corresponds to the ground-state energy of the 1D spin chain and has the value (per unit length) $\Delta E_k/L_y =-2t/\pi$ \cite{PhysRevB.49.14047}. Therefore, the total excitation energy (per unit length) of a domain wall becomes $\Delta E/L_y =\Delta E_p/L_y + \Delta E_k/L_y = (\mu-4V')/2 -2t/\pi$. When its excitation energy is reduced to zero by decreasing $\mu$, the phase transition out of the CBS induced by proliferation of domain walls will occur. This gives the critical $\mu_{c1}$ for the CBS-ICS phase transition, i.e.
\begin{equation}
\mu_{c1}=4V'+\frac{4t}{\pi}.
\label{mu1}
\end{equation}
This quantum phase boundary is plotted in Fig. \ref{fig1} (the upper dotted line), which is in good agreement with the QMC results.

For the transition from QFS to ICS, the same argument applies. Doping the QFS with a column of bosons, it actually generates two domain walls and the total cost in potential energy per unit length equals to $(4V'-\mu)/2$. Here we define the domain wall as the line connecting nn bosons. Again, by mapping each of the two domain walls onto a 1D spin-1/2 chain, the gain in kinetic energy per unit length of the two domain walls is deduced to be $2(-2t/\pi)$. Therefore, the critical $\mu_{c2}$ for the QFS-ICS phase transition is given by
 \begin{equation}
\mu_{c2}=4V'-\frac{8t}{\pi},
  \label{mu2}
\end{equation}
which is again in good agreement with the numerical results. Deviations in analytical predictions of $\mu_{c1}$ and $\mu_{c2}$ are expected for large $t$,  since higher order corrections and SF fluctuations become important then.

\begin{figure}
\includegraphics[width=\columnwidth]{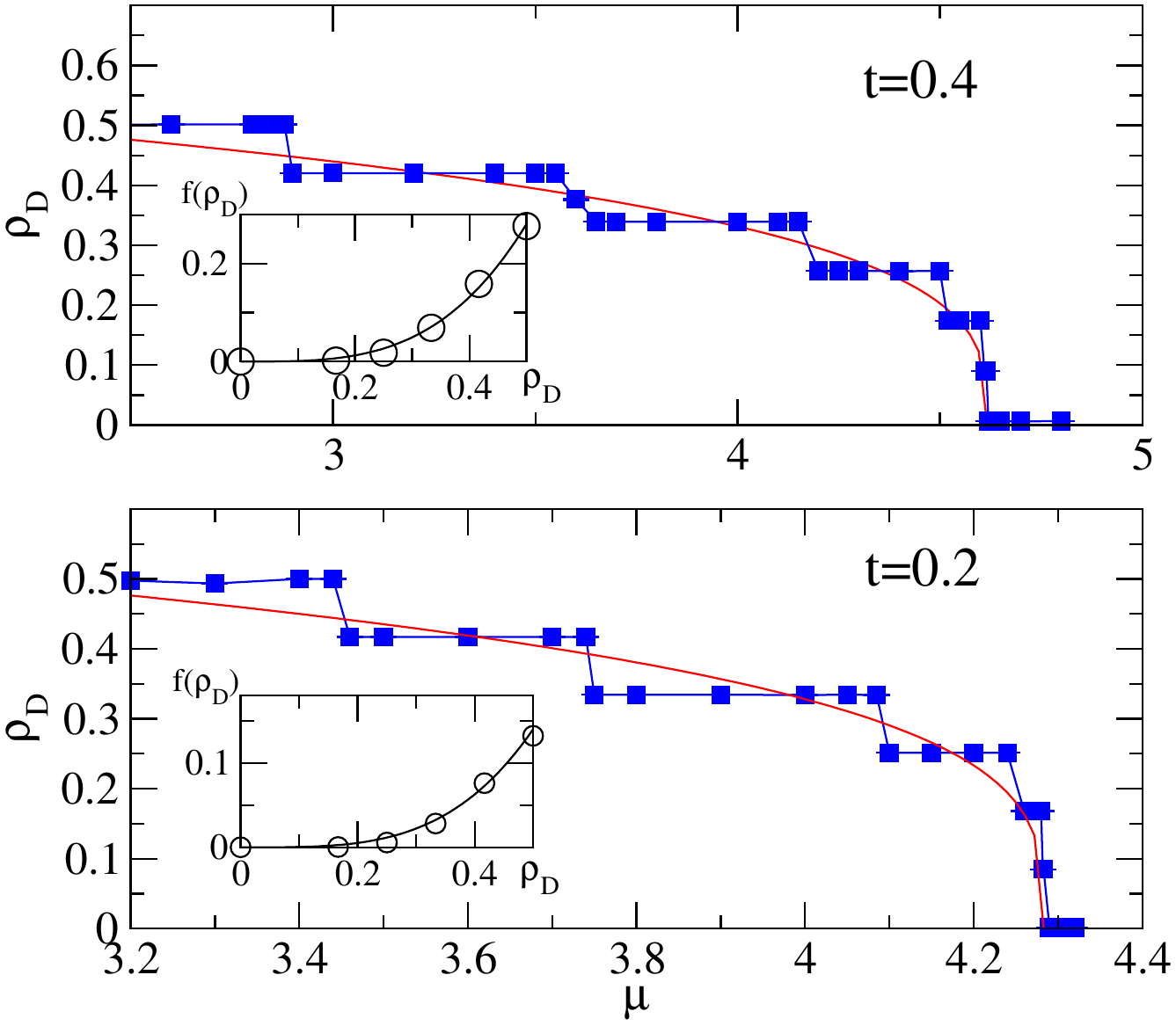}
\caption{Domain wall density $\rho_D$ as a function of $\mu$ for $t=0.4$ (upper panel) and $t=0.2$ (lower panel). The red solid lines are analytical predictions given by Eq.~(\ref{rho_D_2}). The insets show the fittings of the interaction energy $f(\rho_D)$ to a power-law function $A\rho_D^\alpha$ for each case. Here $V=4$ and $L=24$ are taken.
\label{fig4}}
\end{figure}

Following the analysis in Ref.~\cite{PhysRevLett.117.193201}, we also determine the domain wall density $\rho_D$ as a function of $\mu$, which is defined as:
\begin{equation}
\rho_D =\frac{1}{L_x L_y} \sum_{x,y} \bar{n}_{x,y} \bar{n}_{x+1,y} \; ,
  \label{rho_D}
\end{equation}
where $\bar{n}_{x,y}=1-n_{x,y}$ is the hole number at site $(x,y)$. Here the summation of $\bar{n}_{x,y} \bar{n}_{x+1,y}$ over $x$ coordinates counts the number of hole-hole bonds in the $x$ direction, and then gives the number of linear domain walls along the $y$ direction. The QMC results of $\rho_D$ are shown in Fig. \ref{fig4}.

Clearly, the domain walls will interact with each other. In combination with the discussions in the first paragraph of this section for isolated domain walls, the total energy for interacting domain walls at a finite density $\rho_D$ will take the following form:
\begin{equation}
E(\rho_D) = L_x L_y V' \rho_D \left[ -2 + \frac{\mu}{2 V'} - \frac{2 t}{\pi V'} + f(\rho_D) \right].
  \label{E_rho_D}
\end{equation}
Here the last term accounts for an effective repulsive interaction energy between two neighboring domain walls. This yet unknown function $f(\rho_D)$ can be determined by fitting to the numerical results.

We note that the transition from the state with $2M-2$ domain walls to that with $2M$ domain walls occurs when $E[(2M-2)/L_x]=E(2M/L_x)$. This condition leads to a recursive relation of $f(2M/L_x)$:
\begin{equation}
\mu_M =  \mu_{c1} + 2V' \left[ (M-1)f(\frac{2M-2}{L_x}) - M f(\frac{2M}{L_x}) \right] \; ,
  \label{mu_M}
\end{equation}
which can be solved to have
\begin{equation}
f(\frac{2M}{L_x})=\frac{1}{V'} \left[ \frac{\mu_{c1}}{2} - \sum^M_{i=1} \frac{\mu_i}{2M} \right].
  \label{f_2M}
\end{equation}

As a result, the simulated values of $\mu_M$ can be used to evaluate the discrete values of $f(2M/L_x)$. As shown in the inset of Fig. \ref{fig4}, a power-law function $A\rho_D^\alpha$ can be fitted to $f(\rho_D)$ with the exponent $\alpha = 3.4(2)$ for $t=0.4$ (upper panel). With this result, the domain wall energy of Eq. \ref{E_rho_D} is then minimized to obtain the domain wall density
\begin{equation}
\rho_D=\left[ \frac{\mu_{c1}-\mu}{2V'A(\alpha+1)} \right]^{1/\alpha} \; ,
  \label{rho_D_2}
\end{equation}
which is found to be in good agreement with the simulated data in Fig.~\ref{fig4}. Our results thus justify the application of the domain wall dynamics in the present system.
Interestingly, we repeated the same calculations on domain wall density for $t=0.2$ and obtained nearly the same exponent $\alpha=3.6(3)$. This suggests a possible universal exponent for different hopping integrals.

\section{Incommensurate-superfluid and thermal phase transitions}
\label{QPT}
\begin{figure}
\includegraphics[width=7cm]{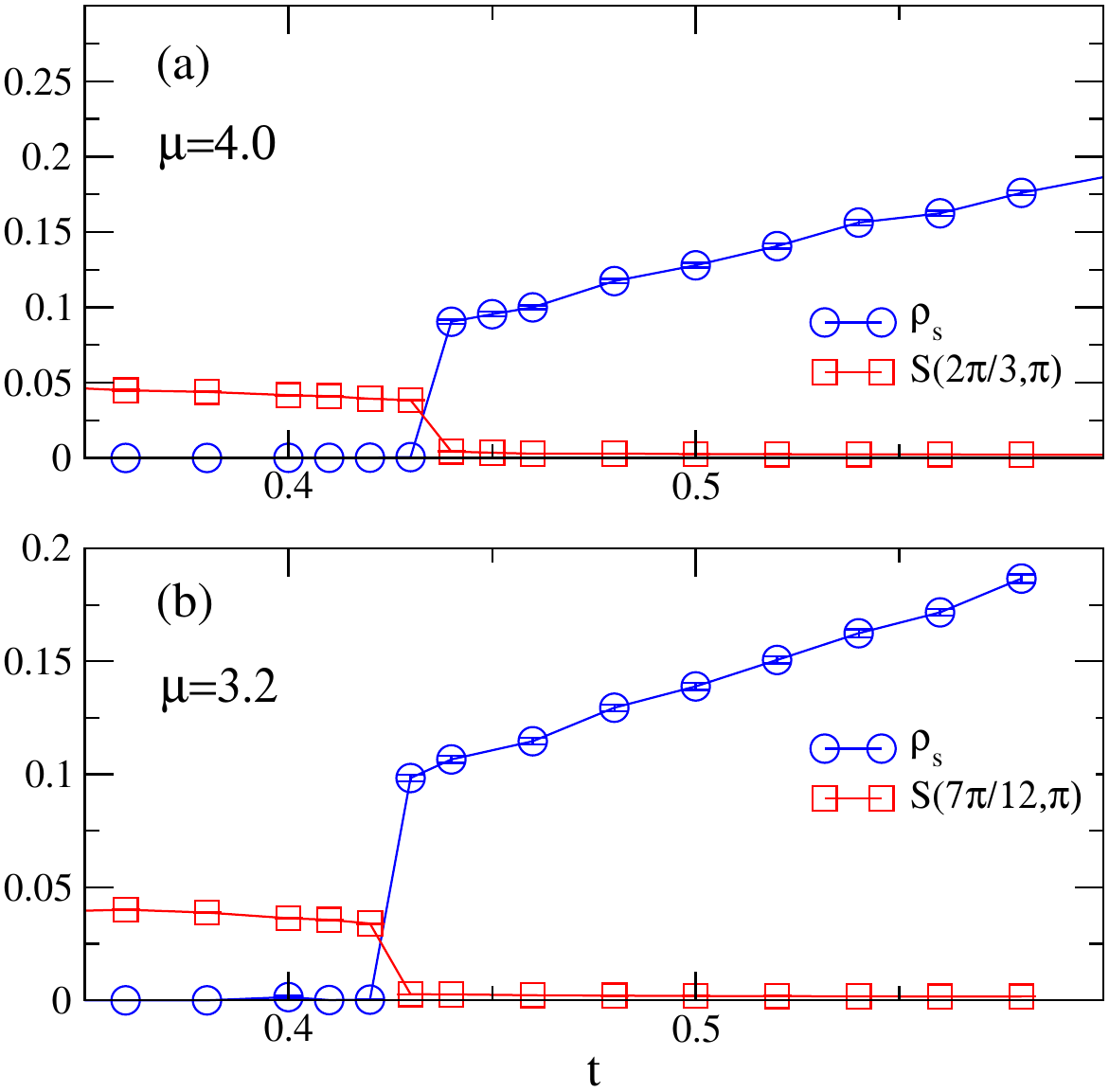}
\caption{Superfluid density $\rho_s$ (open circles) and structure factor $S(\textbf{Q})$ (open squares) as functions of $t$ for (a) $\mu=4.0$ and (b) $\mu=3.2$. Here $V=4$ and $L=24$. Both quantities show the signatures of first-order phase transition from the ICS to the SF states.
\label{fig5}}
\end{figure}

In our parameter regime, the quantum phase transitions from all solid states to the SF phase are observed to be discontinuous and there appears no immediate supersolid phase. These discontinuous transitions are demonstrated by the abrupt changes of the superfluid density $\rho_s$ and the structure factor $S(\textbf{Q})$ at the phase boundaries. Here we focus on the transitions out of the ICS states.

In Fig. \ref{fig5}, we present our data for $\mu=4.0$ and 3.2, which give the ICS states at small hopping $t$ with $\rho=1/3$ and 7/24 respectively. Upon increasing $t$, we find that both $\rho_s$ and $S(\textbf{Q})$ change abruptly across the phase boundaries. We note that the solid ordering is destroyed at the same critical $t$ where the superfluidity emerges. Hence no signs of coexistence of both order parameters is found. The absence of supersolid phase could be accounted for by the same argument for the CBS and the QFS that phase separation is more favorable than the supersolid phase \cite{PhysRevLett.84.1599}.

%For the transitions from the incommensurate solids to the CBS or the QFS, in the thermodynamic limit where there are infinite number of plateau states, the particle density $\rho$ is expected to be varied continuous from 1/4 to 1/2. While the broken symmetry of CBS is different from that of QFS, it is interesting to note that a continuous change of broken symmetry in the ICS connects these two phases via two second-order phase transitions.

\begin{figure}
\includegraphics[width=\columnwidth]{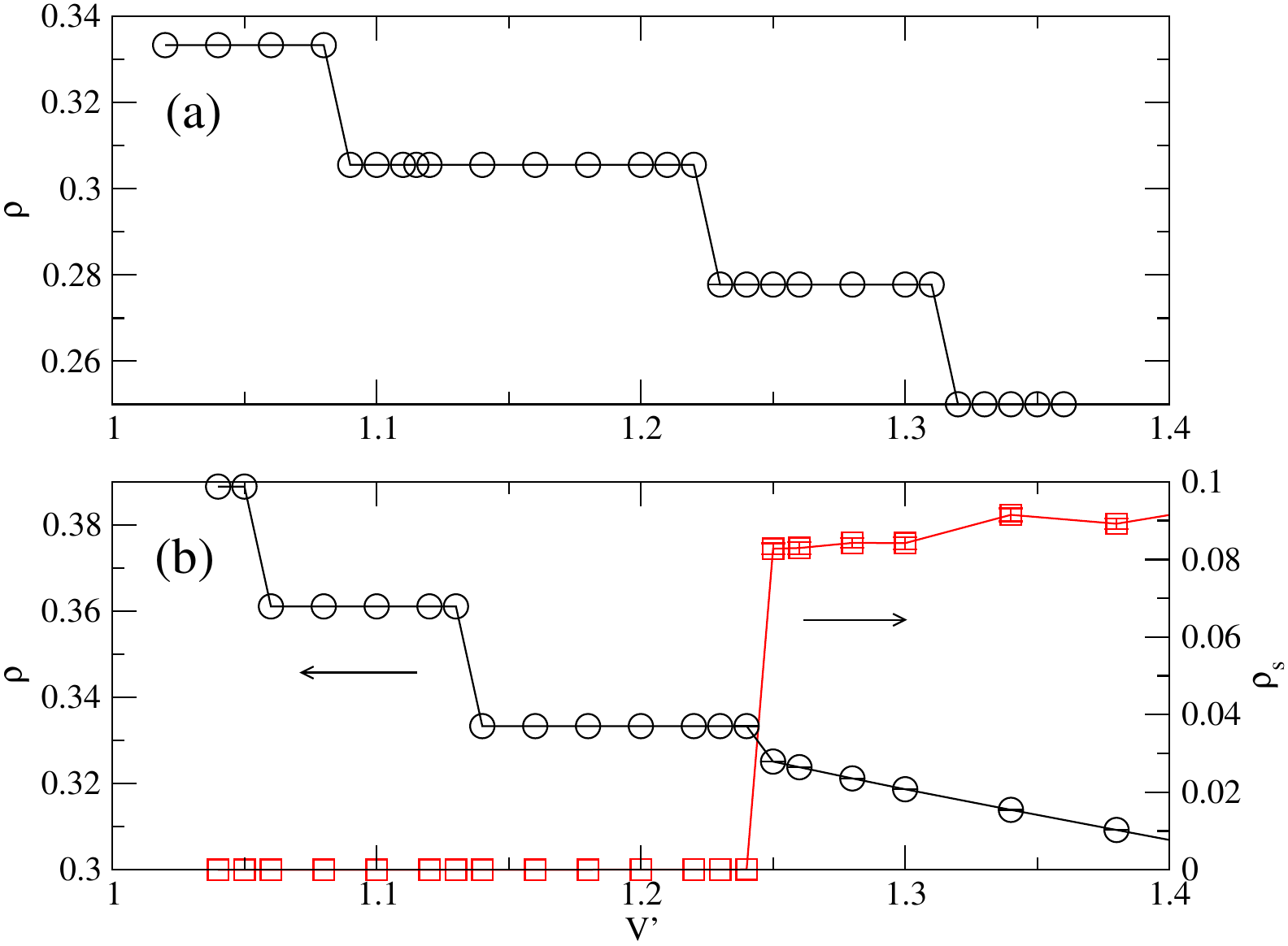}
\caption{Boson density $\rho$ (circles) and superfluid density $\rho_s$ (squares) as functions of $V'$ with $V=4$ for (a) $\mu=4.0$ and (b) $\mu=4.6$. Here $L=36$ and $V=4$.
\label{fig6}}
\end{figure}

In the previous discussions, we fix the nnn interaction $V'$ to be unity. We now turn to the effect from varying $V'$.
When $V'$ is increased, the enhanced nnn frustrations will reduce the cost in potential energy for the domain wall formation (see Sec. \ref{DW}). Therefore, the boson density of the system is expected to follow a series of density drops as more domain walls of holes can be generated for larger $V'$. Our numerical results for $V=4$ at $\mu=4.0$ and 4.6 plotted in Fig. \ref{fig6} demonstrate this prediction. At $\mu=4.0$, the system moves from the 1/3-filling state all the way to the quarter-filling state as $V'$ increases. However, at $\mu=4.6$, when $V'>1.24$, the frustration becomes too strong such that the diagonal long-range order cannot be sustained and the system melts into a superfluid phase instead. An abrupt jump in superfluid density is observed that signals a direct first-order phase transition from the ICS to the SF and there exists no intermediate supersolid phase.

\begin{figure}
\includegraphics[width=\columnwidth]{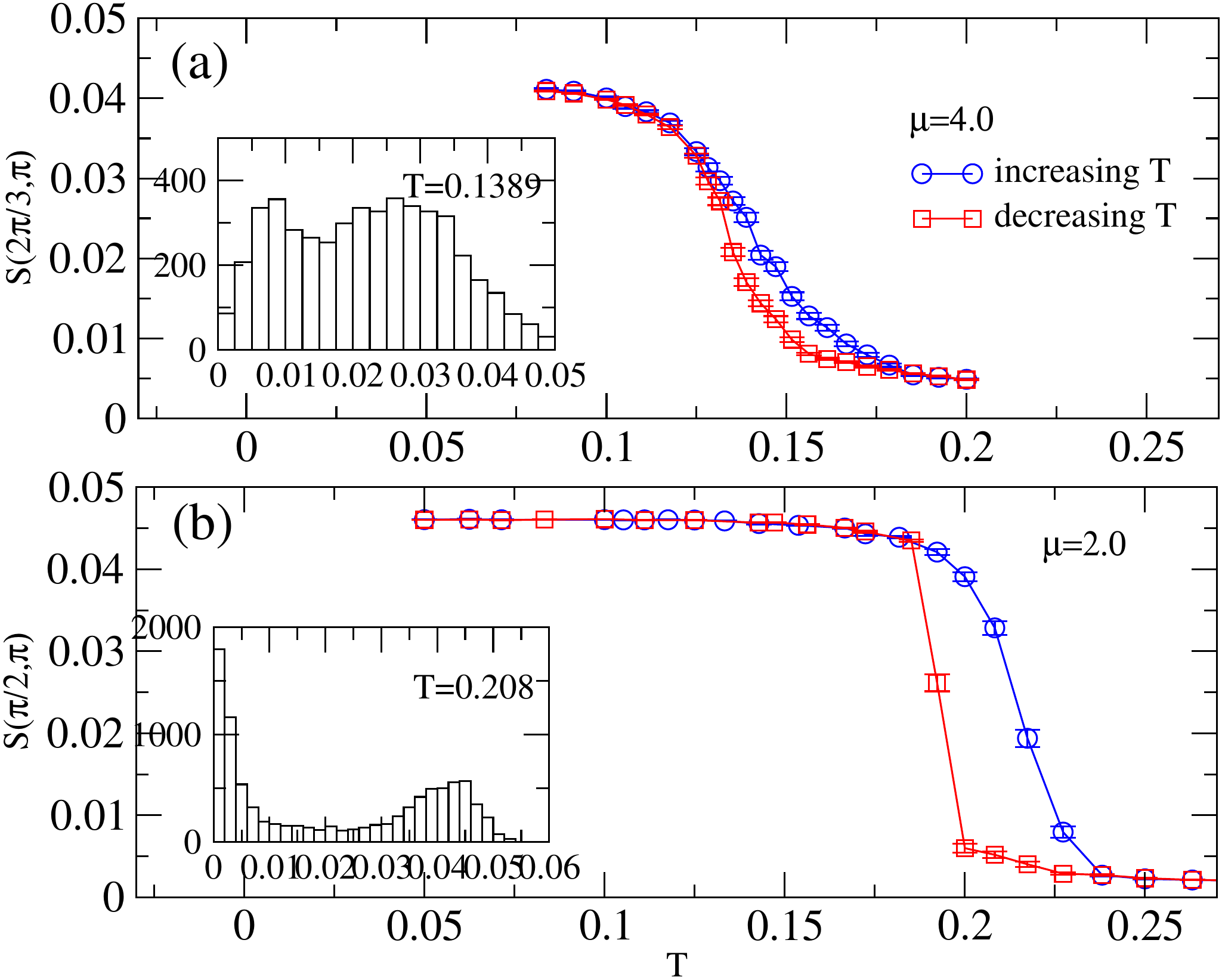}
\caption{Hysteresis of the structure factors S($\textbf{Q}$) for the cases of (a) $\mu=4.0$ and (b) $\mu=2.0$, which correspond to the 1/3-filling plateau state and the 1/4-filling QFS state respectively. Here $V=4$ and $L=24$. The insets show the double-peak feature of S($\textbf{Q}$) around the transitions.
\label{fig7}}
\end{figure}

We now consider the thermal phase transitions out of the ICS and the QFS phases. It is expected that the broken translation symmetry in these two phases will be restored by thermal fluctuations at high temperatures. This symmetry restoration can be detected by the disappearance of the corresponding structure factors. Careful studies of the order-disorder transitions indicate that these transitions are of first order. In Fig. \ref{fig7}, we measure the structure factors $S(2 \pi/3, \pi)$ and $S(\pi/2,\pi)$ of the $\rho=1/3$ ICS state and the $\rho=1/4$ QFS state respectively. At the first glance, the thermal transitions seem to be continuous, but a detailed analysis shows the typical hysteresis behaviors of discontinuous phase transitions around the transition temperatures. Starting from the order (disorder) states, we increase (decrease) the temperature slowly in the QMC simulations by using the operator lists taken from previous temperatures as initial conditions. From Fig. \ref{fig7}, it is clearly that results from increasing and decreasing temperatures follow different paths around the phase transitions. Double peaks are also observed in the histograms (insets in Fig. \ref{fig7}) which again confirm the nature of first-order transitions. We note that the previous study for another type of quarter-filled solid suggested a continuous thermal phase transition \cite{PhysRevB.82.184505}. This difference can be explained by the distinct broken symmetries of these two types of quarter-filled solids. Furthermore, it is noticeable that the hysteresis loop becomes much smaller and the double peaks feature less significantly in the ICS than what happen in the QFS. This indicates that, as the chemical potential $\mu$ increases from the QFS to the ICS, the thermal transition becomes a more weakly first-order one. This is consistent with the fact that the thermal transition of half-filled CBS is known to be continuous, so that approaching the CBS from the QFS via the ICS, the thermal transitions are expected to evolve from strongly first-order to weakly first-order and then to second-order. Since the ordering wave vector $\mathbf{Q}$ of the maximal structure factor varies continuously in the ICS phase into the CBS, it is natural to expect a smooth evolution of transition order instead of an abrupt change as $\mathbf{Q}$ approaches to ($\pi$, $\pi$).

\section{Conclusion}
\label{CN}
We have shown that the ICS states can emerge in a bosonic model on a square lattice under short-range frustrated interactions. This observation has been overlooked in previous studies of the extended Bose-Hubbard model. The ICS phase appears in between the half-filled and the quarter-filled solids and is characterized by a series of density plateaus with fractional values of densities. The short-range character of the interactions in our model simplifies the analysis and allows us to address fundamental questions about the nature of the plateau states. By following the analysis in Ref. \cite{PhysRevLett.117.193201}, we show that the fractional particle densities and the ordering wave vectors of the plateau states can be explained by the domain wall formation. This indicates that the incommensurate phase originates from proliferation of domain walls. Furthermore, the measured domain wall densities $\rho_D$ agree well with the predicted values from the theory of the interacting domain walls. We find that the interaction term behaves as a power-law function of $\rho_D$ with an exponent likely to be independent of the system parapmeters. In the thermodynamic limit, we expect the ordering wave vector $\mathbf{Q}$ characterizing different broken symmetries changes continuously from $\mathbf{Q}=(\pi/2, \pi)$ for the quarter-filled star order to $\mathbf{Q}=(\pi, \pi)$ for the checkerboard order. In addition, the widths of the intermediate plateau states become diminished. There is no superfluidity found in the ICS phase, which differs from the case of anisotropic triangular lattice where an incommensurate supersolid is observed \cite{PhysRevLett.117.193201}.
It is not surprising since, in contrast to the case of the square lattice, bosons have more degrees of freedom to hop around on the triangular lattice. Based on this observation, incommensurate supersolids may come out in our model if the nnn hopping is turned on. It is also interesting to consider the effect of strong frustration (large $ V'$) on the ICS phase. We show that, by increasing $V'$, either the ICS state is melted into a SF or the system is transformed to be a QFS. Nevertheless, it is known that for even stronger $V'$ another type of star order becomes favorable at quarter filling. Whether similar ICS phases can still appear around such quarter-filled solids is not clear without extensive studies on a wide range of parameter regime. Furthermore, one may wonder if the present mechanism of domain wall formation may be applied to explain the results observed in the infinite-range models. This is worth investigating in a further study.

\begin{acknowledgments}
The numerical computations are performed in the Center for High Performance Computing of the THU.
This work is financial supported by the Ministry of Science and Technology of Taiwan under Grant No. MOST 108-2112-M-029 -005 and MOST 109-2112-M-029-006. M.F.Y. is also supported by the Ministry of Science and Technology of Taiwan under Grant No. MOST 109-2112-M-029-005.
\end{acknowledgments}

%apsrev4-2.bst 2019-01-14 (MD) hand-edited version of apsrev4-1.bst
%Control: key (0)
%Control: author (8) initials jnrlst
%Control: editor formatted (1) identically to author
%Control: production of article title (0) allowed
%Control: page (0) single
%Control: year (1) truncated
%Control: production of eprint (0) enabled
%

%\bibliography{xBH}
\end{document}